\documentclass{PoS}

\title{The isospin breaking effect on baryons with $N_f=2$ domain wall fermions}

\ShortTitle{The isospin breaking effect on baryons with $N_f=2$ domain wall fermions}

\author{
   \speaker{T. Doi},$^a$\thanks{Email: doi@quark.phy.bnl.gov.}
   T. Blum,$^{a,b}$ M. Hayakawa,$^c$ 
   T. Izubuchi$^{a,d}$ 
   and
   N. Yamada$^{e,f}$\\
\llap{$^a$}RIKEN BNL Research Center, Brookhaven National Laboratory, Upton, NY 11973, USA\\
\llap{$^b$}Physics Department, University of Connecticut, Storrs, CT 06269, USA\\
\llap{$^c$}Department of Physics, Nagoya University, Nagoya 464-8602, Japan\\
\llap{$^d$}Institute for Theoretical Physics, Kanazawa University, Kakuma, Kanazawa 920-1192, Japan\\
\llap{$^e$}High Energy Accelerator Research Organization (KEK), Tsukuba 305-0801,Japan\\
\llap{$^f$}School of High Energy Accelerator Science, The Graduate University for Advanced Studies (Sokendai), Tsukuba 305-0801, Japan\\

{\rm\bf for the RBC collaboration}

}

\abstract{
We study the isospin breaking effect on octet baryons.
Using the two-flavor dynamical domain-wall QCD configurations 
combined with the quenched non-compact QED configurations,
the electromagnetic mass splittings between isomultiplets 
$(p, n), (\Sigp, \Sigz, \Sigm), (\Xiz, \Xim)$
are investigated.
We evaluate the main source of statistical fluctuations in the 
two-point correlation function,
and find that the elimination of ${\cal O}(e)$ fluctuation  ($e$: the QED charge)
is essential to extract the signal.
Preliminary results for $m_p - m_n$ as well as other mass splittings are presented.
Possible origin of systematic uncertainty is also discussed.

}

\FullConference{XXIVth International Symposium on Lattice Field Theory\\
                July 23-28, 2006\\
                Tucson, Arizona, USA}

\makeatletter
\def\simleq{\mathrel{\mathpalette\gl@align<}}
\def\simgeq{\mathrel{\mathpalette\gl@align>}}
\def\gl@align#1#2{\lower.6ex\vbox{\baselineskip\z@skip\lineskip\z@
     \ialign{$\m@th#1\hfill##\hfil$\crcr#2\crcr\sim\crcr}}}
\makeatother

\newcommand{\bra}{\langle}
\newcommand{\ket}{\rangle}
\newcommand{\braket}[1]{\bra #1 \ket}
\newcommand{\msea}{m_{\rm sea}}
\newcommand{\mval}{m_{\rm valence}}
\newcommand{\Uqcd}{U_\mu^{\rm QCD}(x)}
\newcommand{\Uqed}{U_\mu^{\rm QED}(x)}
\newcommand{\Aqed}{A_\mu^{\rm QED}}
\newcommand{\alem}{\alpha_{\rm em}}
\newcommand{\Sigz}{\Sigma^0}
\newcommand{\Sigp}{\Sigma^+}
\newcommand{\Sigm}{\Sigma^-}

\newcommand{\mSigall}{m_{\Sigp} + m_{\Sigm} - 2 m_{\Sigz}}
\newcommand{\Xiz}{\Xi^0}
\newcommand{\Xim}{\Xi^-}
\newcommand{\Lam}{\Lambda}

\begin{document}

\section{Introduction}

Recent development of the lattice QCD simulation 
enables us to access hadron properties with great accuracy.
In fact, as a precise first principle calculation of QCD,
the result of lattice QCD is now being used for examination
of the fundamental theory, such as a unitarity triangle test
of the standard model\cite{onogi-lat06}.
Yet, there are various quantities remained unsettled so far.
Among them, we focus on the isospin breaking effect on baryons,
such as proton-neutron ($p-n$) mass difference.
Actually, this mass difference is one of the most fundamental quantities
in nuclear physics.
For instance,
this quantity governs the $\beta$-decay of the neutron, i.e., 
the life time of the neutron.
Note also that $\beta$-decay/electron capture
are the basic ingredients
in the understanding of the nucleosynthesis 
and the history of the universe:
if the mass ordering between proton and neutron
was opposite, the universe may not exist 
as it is now.
In the laboratory-experiment,
the isospin breaking effect on the baryon spectrum
is observed not only for $p-n$ 
mass splitting 
but also for splitting between other isomultiplets in octet/decuplet members.
The charge symmetry breaking in the $N-N$ interaction
is also observed, which is caused by the isospin breaking of nucleons.

The isospin breaking of hadrons originates in two ingredients:
one is the up and down quark mass difference in QCD, 
and the other is the electromagnetic (EM) effect based on QED.
In this sense,
the determination of u, d quark mass through the 
study of isospin breaking corresponds to
fixing the Yukawa coupling constant between
u, d quark and the Higgs particle,
which are the fundamental (and unknown {\it a priori})
parameters in the standard model.
Moreover, u, d quark masses are particularly interesting 
from the viewpoint of the so-called strong CP problem.
In fact, it is proposed that the existence of 
massless quark(s) can absorb the phase $\theta$
of the vacuum, and thus resolve the strong CP problem.
By studying the isospin breaking in hadron spectrum,
we can examine whether or not such a scenario 
happens in our realistic world.
On the other hand, the inclusion of QED in 
the QCD calculation is becoming the urgent
task to develop the new physics search
beyond the standard model.
For instance, in the theoretical calculation
of muon anomalous magnetic moment,
it is know that hadronic light-by-light scattering contribution
makes the large uncertainty,
and it is proposed to resolve this problem
using the QCD + QED simulation\cite{masashi-lat05}.
From this viewpoint, 
the study of isospin breaking on the spectrum
is a suitable topic to build a firm foundation
for the QCD + QED simulation.

In the literature of lattice QCD, the first work
has been done by Duncan {\it et al.}\cite{duncan}, where they employed
the quenched non-compact QED combined with quenched QCD 
with the Wilson fermion.
In this pioneering work, however,
both of the QCD quenching artifact and the 
artificial chiral symmetry breaking contamination
are unavoidable.
%
%
In this work,
we eliminate 
these uncertainties by 
employing $N_f=2$ dynamical domain-wall QCD configuration.
Our study for the meson sector (such as $\pi^+ - \pi^0$, $K^+ - K^0$ mass splitting)
has been already reported in Ref.\cite{nori-lat05},
and we report the study of the EM effect on octet baryons in this proceeding.
The effect of u, d quark mass difference will be reported elsewhere\cite{doi-future}.
There is another work\cite{namekawa-lat05} on splitting between meson isomultiplets
with improved action but still at the quenched level.
Recently, the splitting of $p - n$ is also studied\cite{savage-pn},
while only the effect of u, d quark mass difference is considered there.

\section{Formalism}

We study the EM effect on baryons through
the mass difference between isomultiplets, such as
$m_p-m_n$, 
$m_{\Sigma^+} - m_{\Sigma^0}$,
$m_{\Sigma^-} - m_{\Sigma^0}$,
$m_{\Sigma^+} - m_{\Sigma^-}$,
$m_{\Sigma^+} + m_{\Sigma^-} - 2 m_{\Sigma^0}$,
$m_{\Xi^-} - m_{\Xi^0}$
using the QCD + QED simulation.
In the QCD sector, we employ the $N_f=2$ unquenched QCD gauge
configuration $\Uqcd$ generated by the RBC collaboration\cite{rbc-2f}.
The domain-wall fermion action and the DBW2 gauge action
is used with the parameters of 
$V= L^3 \times T = 16^3\times 32$, $L_s=12$, $M_5=1.8, \beta=0.8$.
The ensembles are generated for three different (u,d-) sea quark 
masses of $\msea = 0.02, 0.03, 0.04$.
The lattice unit is determined to be $1.691(53)$ GeV so as to 
reproduce $\rho$ meson mass $m_\rho=770$MeV.
The sea quark mass and the physical volume roughly correspond to 
$\msea \simeq \frac{1}{2} m_s - m_s (m_s: \mbox{strange quark mass})$ 
and $L^3 \simeq (1.9{\rm fm})^3$, respectively.
We pick up about 200 configurations from $\sim 5000$ trajectories
available at each sea quark mass ensemble.
The analysis is performed with bin size of 2,
(i.e., bin size of 50 trajectory separation),
in order to suppress the possible autocorrelation.


In the QED sector, we employ a non-compact formulation of \\
\noindent
$
{\cal L}_{\rm QED} = \frac{1}{4} 
\sum_{x, \mu, \nu} 
\left(\partial_\mu {A_\nu^{\rm QED}}(x) - \partial_\nu {A_\mu^{\rm QED}}(x)\right)^2$
at the quenched level\cite{duncan}.
In the generation,
we first generate $\Aqed$ in the momentum space 
under the Coulomb plus residual gauge fixing condition together with
a boundary condition for the constant modes.
The configuration in the coordinate space is obtained
by Fourier transformation.
The advantage of this formulation is that
the generation of $\Aqed$ leads to just a 
Gaussian random number generation,
and thus there is no autocorrelation
between the configuration even for 
arbitrary small coupling.
In addition to that,
we do not have to worry
about the renormalization of the QED coupling constant,
because 
the quenched QED in non-compact formulation
is a free theory.
Given the QED configuration described above,
we obtain a $U(1)$ link variable by
$\Uqed = \exp[-i\Aqed(x)]$, and then 
construct the QCD + QED configuration
as $\Uqcd \times (\Uqed)^Q$, where $Q = +2/3 e , -1/3 e$
for u and d quark, respectively.
In order to study the QED charge dependence of the mass splitting,
we use not only the physical QED charge, 
$\alem \equiv e^2/4\pi = \frac{1}{137} (\equiv \alem^{\rm phy})$,
but also the charges of 
$\alem = (0), 
\frac{(0.6)^2}{4\pi}, \frac{(0.85)^2}{4\pi}, \frac{(1.0)^2}{4\pi}$
(except for $\alem = \frac{(0.85)^2}{4\pi}$ at $\msea=0.02$).

The mass of baryon $B$ is measured using the two-point correlation function\\
\noindent
$
\Pi_{BB}(t) = \sum_{\vec{x}} \braket{J_B(x) \bar{J}_B(0)}$
with the use of positive parity projection.
We use the operator $J_B$ which has non-relativistic limit,
for instance, 
$J_p = \epsilon_{abc} (u_a^T C \gamma_5 d_b) u_c$ as the proton operator.
The operators for other octets and singlet are obtained by the
SU(3) rotation.
Although these procedures are valid for $p, n, \Sigma^+, \Sigma^-, \Xi^-, \Xi^0$
baryons, additional treatment is necessary 
in the $Q=0, S=-1$ channel, i.e., for $\Sigma^0, \Lambda_8, \Lambda_1$ baryons.
In fact, 
the mixing between $\Lambda_8$ and $\Lambda_1$ occurs 
because of the SU(3) breaking,
and the mixing between $\Sigma^0, \Lambda_8, \Lambda_1$ occurs
because of the SU(2) breaking.
Note here that
$\Sigma^0(1193)$ is massive than $\Lambda(1116)$, experimentally.
Therefore, in order to determine the mass splitting such as $\mSigall$,
we have to extract the $\Sigma^0$ state 
as a first excited state in $Q=0, S=-1$ channel.
For this purpose, we employ the so-called 
variational method\cite{variational}.
In practice, 
we calculate not only the diagonal correlation functions as 
$\Pi_{B_f B_i}(t) = \sum_{\vec{x}} \braket{J_{B_f}(x) \bar{J}_{B_i}(0)}, 
B_i=B_f \in \{\Sigma^0, \Lambda_8, \Lambda_1\}$
but also the off-diagonal correlation functions as 
$\Pi_{B_f B_i}(t), B_i \neq B_f \in \{\Sigma^0, \Lambda_8, \Lambda_1\}$.
By performing the diagonalization of the $3\times 3$ correlation function matrix
of $\Pi(t=t_0)^{-1} \cdot \Pi(t)$,
we can extract not only the ground state $\Lambda$ but also
the first excited state $\Sigma^0$.

\section{Numerical results and discussions}

Using $N_f=2$ unquenched domain-wall QCD configuration
combined with the quenched QED configuration,
we calculate the baryon two-point correlation functions
for all octet and singlet baryons
including the off-diagonal correlation functions
in $Q=0, S=-1$ channel.
By solving the inverse of the Dirac operator
under the QCD + QED configuration, the EM effect
is automatically taken into account for the valence quark.
In order to suppress the contamination from excited states,
we employ the wall source and point sink correlation function
under the Coulomb gauge fixing.
In this study, we consider the isospin breaking effect 
up to the first order.
Therefore, in the evaluation of the EM effect,
we can use the quark masses without the EM effect.
Namely, 
we perform the lattice simulation only at the unitarity point,
$m_{u, \rm valence} = m_{d, \rm valence}= \msea = 0.02, 0.03, 0.04$
for u, d quarks,
and we use $m_s = 0.0446$ for s quark, 
which is determined from $K$-input\cite{rbc-2f}.

In the $Q=0, S=-1$ channel,
we use the variational method to extract 
the first excited state $\Sigma^0$ as well as
the ground state $\Lambda$.
As described previously, 
we first calculate $3\times 3$ correlation matrix $\Pi(t)$
using the flavor bases of $\Sigz, \Lam_8, \Lam_1$,
and then diagonalize $\Pi(t=t_0)^{-1} \cdot \Pi(t)$,
where the extracted eigenvalues correspond to 
the exponentially decaying correlation function 
for each ground/excited state.
In the following analysis, we fix the arbitrary 
parameter $t_0$ as $t_0=1$.
The dependence on $t_0$ is discussed later.
In the study of the mass splitting between $\Sigma$ triplets,
we perform the similar procedure for $\Sigp, \Sigm$ as well,
i.e., we use $\Pi_{\Sigma\Sigma}(t)/\Pi_{\Sigma\Sigma}(t=t_0)$ 
instead of simple $\Pi_{\Sigma\Sigma}(t)$ where $\Sigma = \Sigp, \Sigm$.
By employing this procedure, we can take the full advantage of the 
statistical correlation among $\Sigma$ triplets
and obtain the reasonable signal.

In order to study the QED charge dependence of the mass splitting, 
we evaluate the QED charges of 
$\alem = (0), \alem^{\rm phy}, \frac{(0.6)^2}{4\pi}, \frac{(0.85)^2}{4\pi}, \frac{(1.0)^2}{4\pi}$
(except for $\alem = \frac{(0.85)^2}{4\pi}$ at $\msea=0.02$).
For each $\alem$, we calculate the correlation function for 
not only $e = +\sqrt{4\pi\alem}$ but also $e = -\sqrt{4\pi\alem}$
and take the average between them.
This corresponds to the use of the QED configuration of 
$\{ \Aqed\}$ $\rightarrow$ $\{ \Aqed, -\Aqed \}$ 
with binning between $\Aqed$ and $-\Aqed$.
In this procedure, ${\cal O}(e)$ contamination in the correlation function
can be eliminated {\it a priori}.
In fact, because the EM effect on physical observables
appears only from ${\cal O}(e^2)$, such ${\cal O}(e)$
contamination is nothing but a statistical noise.
Practically, we find that this procedure actually improve
the S/N drastically, and is essential for the study of the baryon EM splitting.

\begin{figure}[t]
\begin{center}
\begin{tabular}{cc}
\begin{minipage}{73mm}
\begin{center}
\includegraphics[width=\textwidth]{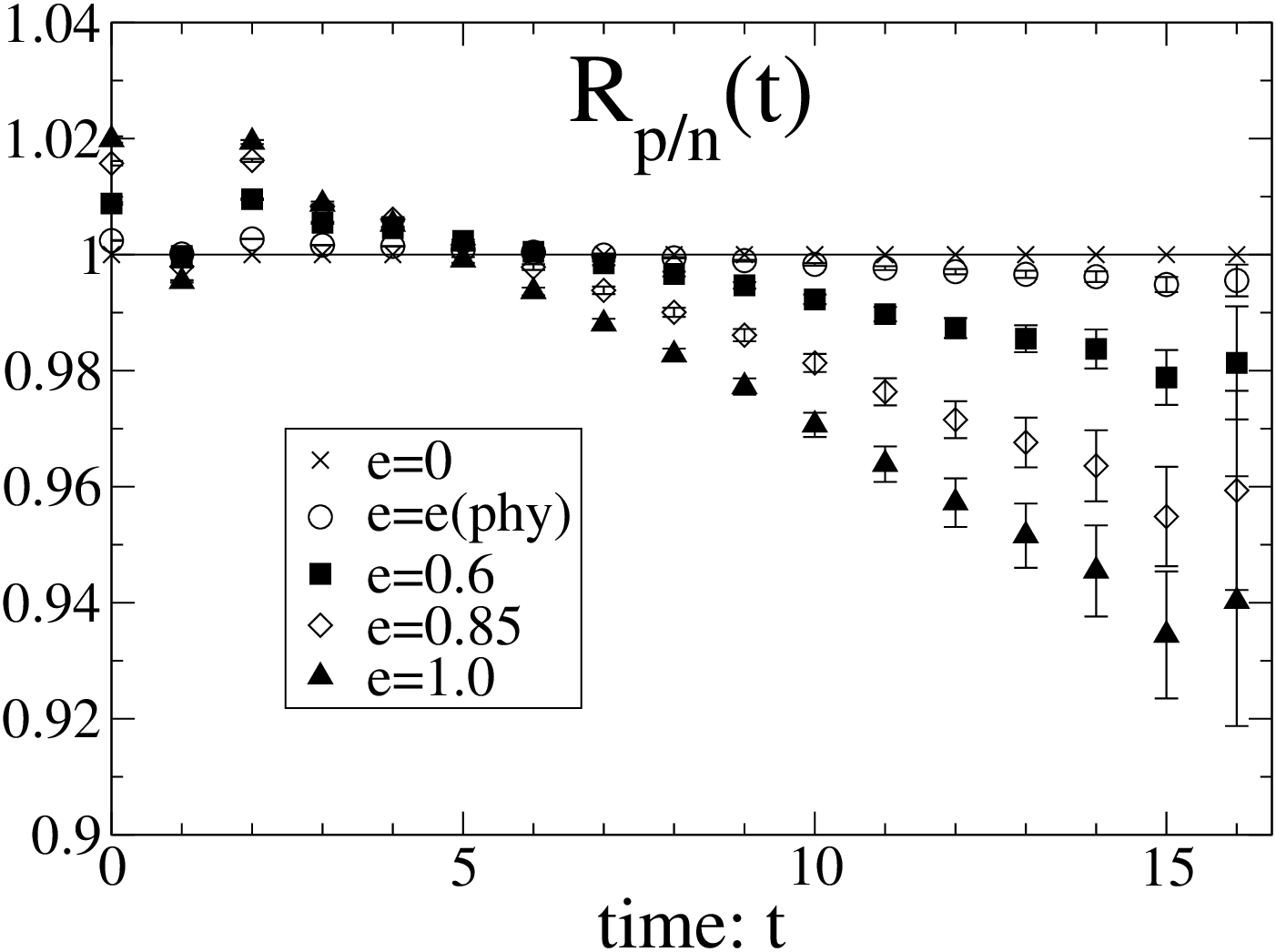}
\caption{
The ratio of the correlation function of proton and neutron $R_{p/n}(t)$ at $\msea=0.03$
is plotted against $t$ for each 
QED charge of $e=(0), e_{\rm phy}, 0.6, 0.85, 1.0$.
Negative linear slope corresponds to $m_p > m_n$ from the EM effect.
\label{fig:PNratio}
}
\end{center}
\end{minipage}&
\begin{minipage}{73mm}
\begin{center}
\hspace*{-3mm}
\includegraphics[width=\textwidth]{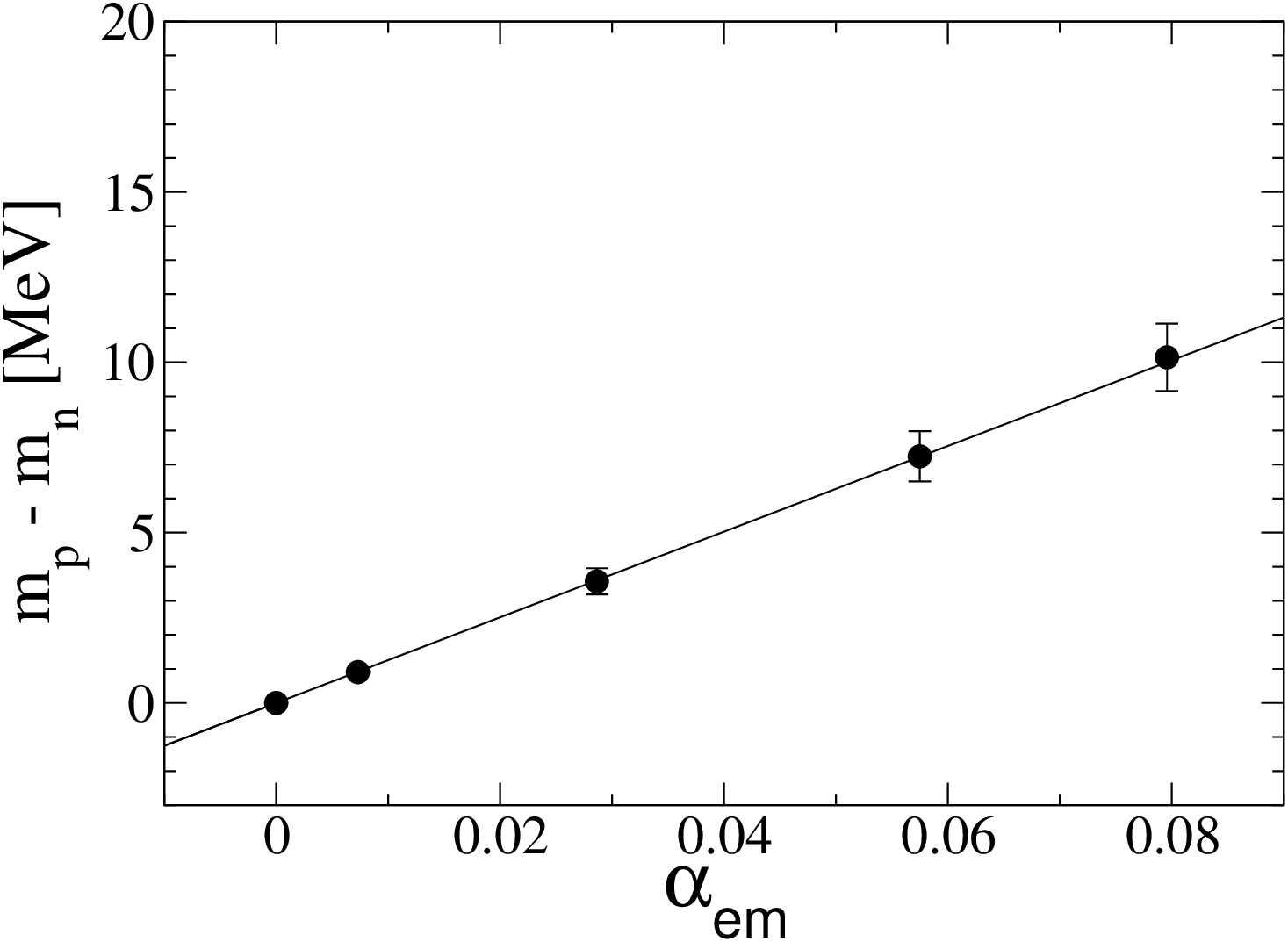}
\caption{The proton-neutron mass difference $m_p - m_n$ from the EM effect
at $\msea=0.03$ is plotted against $\alem\equiv e^2/(4\pi)$. 
The result at $\alem^{\rm phy}$ is shown as the second point from the left.
The solid line corresponds to the best linear fit in terms of $\alem$.
\label{fig:PN_expfit_diff}
}
\end{center}
\end{minipage}
\end{tabular}
\end{center}
\end{figure}

In order to demonstrate the signal of the mass splitting,
we analyze the ratio between the correlation functions of the isomultiplets.
Considering the $p-n$ mass difference for example,
we can express the correlation function for proton and neutron as
$\Pi_{pp}(t) = \lambda_p \exp[-m_p t]$,
$\Pi_{nn}(t) = \lambda_n \exp[-m_n t]$, respectively,
where $\lambda_p$ ($\lambda_n$) is
the proton (neutron) overlap constant between the state and the operator.
Noting that 
$(\lambda_p - \lambda_n)$ and $(m_p-m_n)$ are ${\cal O}(e^2)$, 
we can write the ratio of the proton and neutron correlation functions as
\begin{eqnarray}
R_{p/n}(t) \equiv \Pi_{pp}(t)/\Pi_{nn}(t) 
= 1+ 2 (\lambda_p - \lambda_n)/(\lambda_p + \lambda_n) - (m_p-m_n)\cdot t + {\cal O}(e^4).
\end{eqnarray}
Therefore, the slope of $R_{p/n}(t)$ in terms of the Euclidian time $t$
directly corresponds to the $p - n$ mass splitting.
In Fig.~\ref{fig:PNratio}, we plot $R_{p/n}(t)$ in terms of $t$.
For each QED charge, we find a clear negative linear slope,
which indicates $m_p > m_n$ from the EM effect.



In the practical calculation of the mass splitting, 
we perform the exponential fit for each baryon correlation function
and evaluate the mass difference,
where the statistical error is estimated by the jackknife method.
In Fig.~\ref{fig:PN_expfit_diff}, we plot the $p-n$ splitting
determined at each QED charge $\alem$.
We observe that the splitting behaves linearly in terms of $\alem$,
and there is no indication of the appearance of 
higher order EM effect, ${\cal O}(\alem^2)$.
Because we consider the EM effect up to ${\cal O}(\alem)$ as a framework,
this observation guarantees that our procedure is self-consistent.

\begin{figure}[t]
\begin{center}
\begin{tabular}{cc}
\begin{minipage}{73mm}
\vspace*{-4mm}
\begin{center}
\includegraphics[width=75mm]{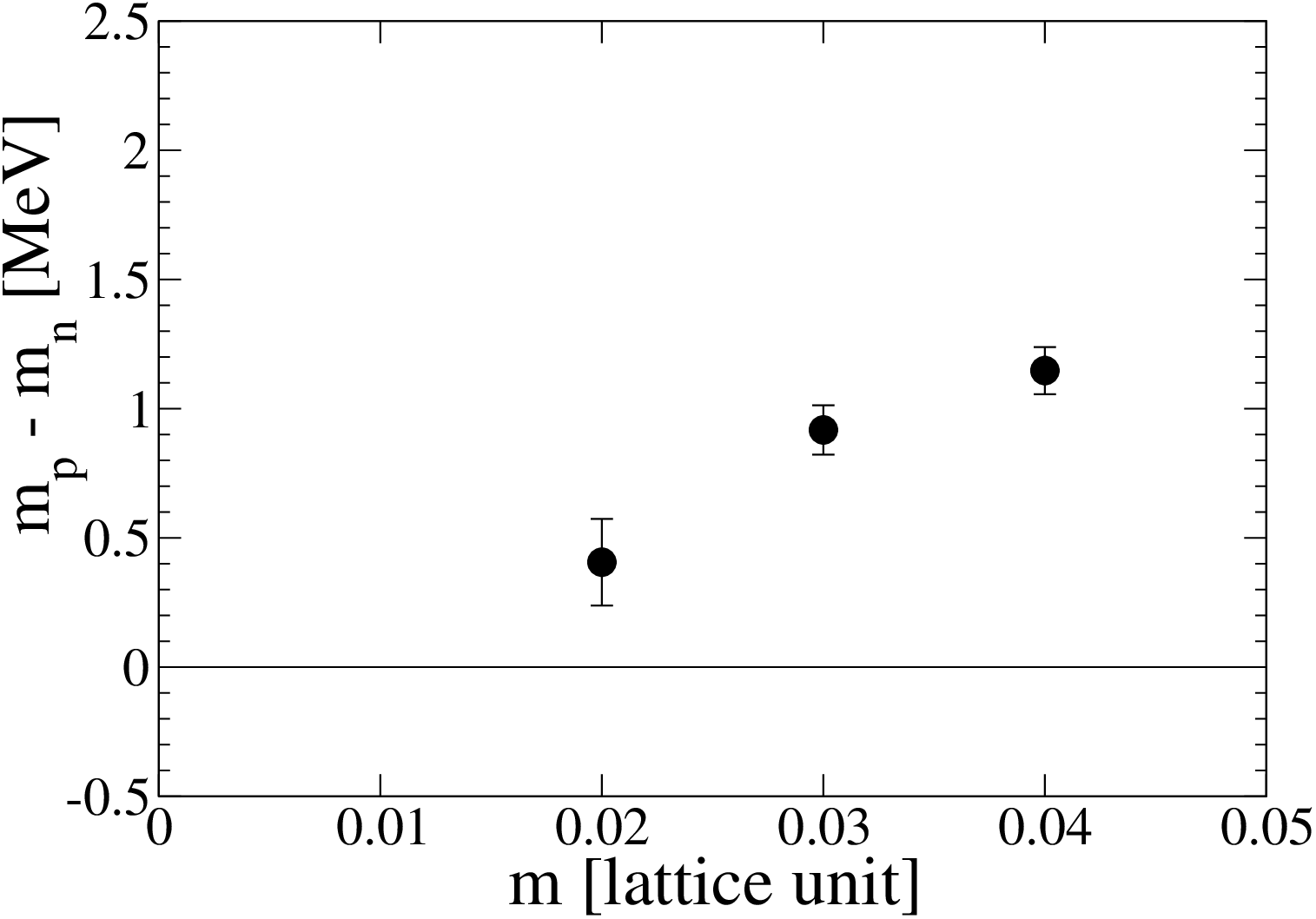}
\caption{The proton-neutron mass difference from the EM effect
at physical QED charge $\alem^{\rm phy}=1/137$ plotted at each quark mass of $m=\mval=\msea$.
\label{fig:all_msea.pn_diff}
}
\end{center}
\end{minipage}&
\begin{minipage}{73mm}
\begin{center}
\hspace*{-3mm}
\includegraphics[width=75mm]{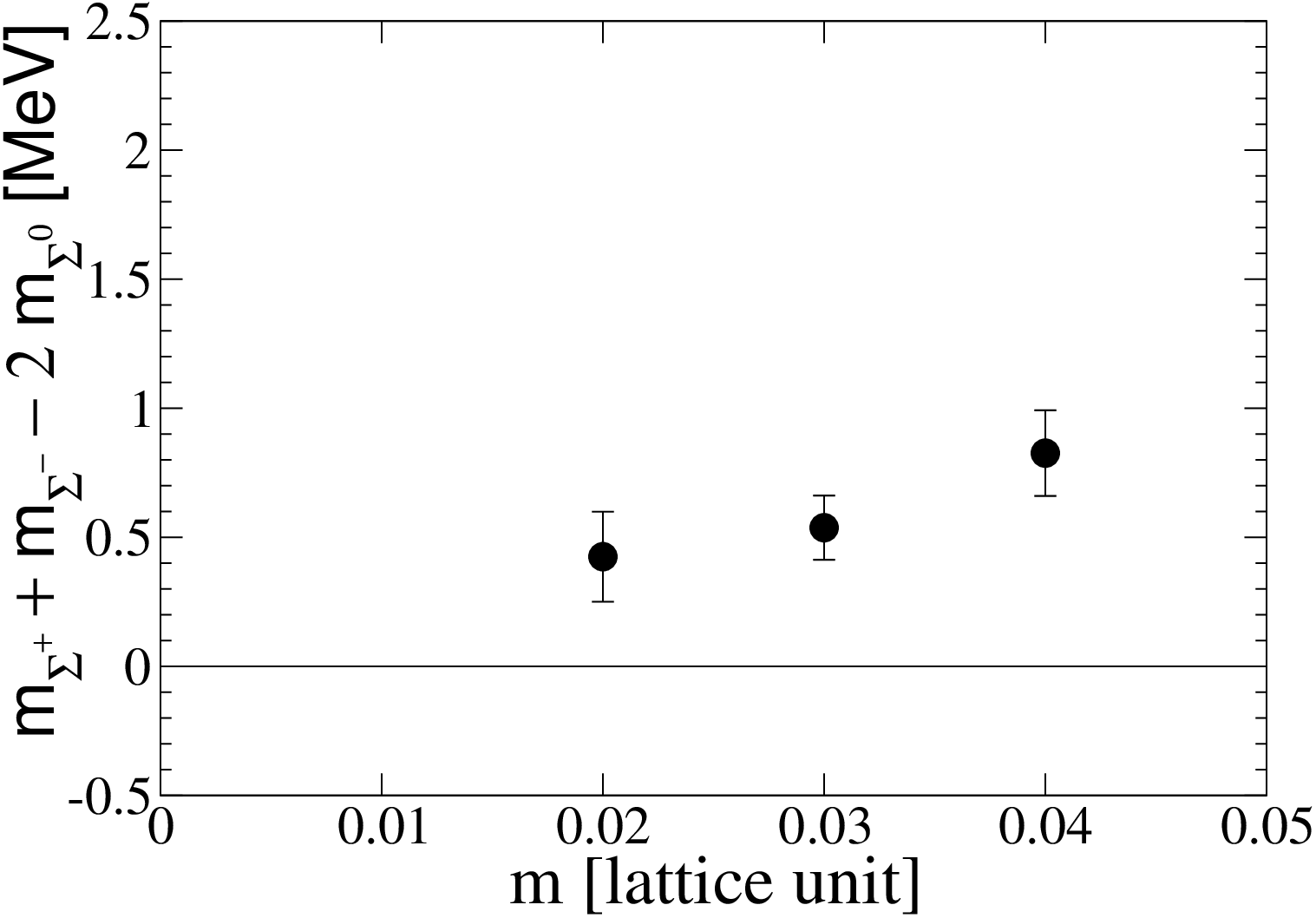}
\caption{The mass difference $\mSigall$
from the EM effect
at physical QED charge $\alem^{\rm phy}$ 
plotted at each quark mass of $m=\mval=\msea$.
\label{fig:all_msea.Sig-all_diff}
}
\end{center}
\end{minipage}
\end{tabular}
\end{center}
\end{figure}

Finally, we fit linearly the splitting in terms of $\alem$,
and determine the splitting at $\alem=\alem^{\rm phy}$.
We perform this procedure at each lattice simulation 
with quark mass of $m = \mval=\msea = 0.02, 0.03, 0.04$.
Fig.~\ref{fig:all_msea.pn_diff} shows the result for 
$p-n$ mass splitting from the EM effect.
We observe the non-trivial EM effect on $p - n$
for each $m$.
Note that these are the first results obtained non-perturbatively
using dynamical lattice simulation.
It is interesting to see that the result at each $m$ is 
roughly consistent with the model estimation
using Cottingham formula, $m_p - m_n = 0.76(30)$MeV\cite{gasser-leutwyler}.
The splitting in the real world can be obtained 
by the chiral extrapolation of the lattice data.
We, however, find that the result at $m=0.02$ is afflicted 
with larger statistical fluctuation, and the reliable 
chiral extrapolation becomes difficult.
In fact, the linear chiral extrapolation in terms of $m$ 
leads to only zero-consistent result for $p-n$ mass difference.
In order to extract better signal,
the statistical improvement is in progress.


In Fig.~\ref{fig:all_msea.Sig-all_diff},
we also show the result of the EM splitting for $\mSigall$ 
at each $m = 0.02$, $0.03$, $0.04$.
The splitting $\mSigall$ is an interesting quantity
from the viewpoint of discrimination 
of the two ingredients in isospin breaking, i.e.,
the EM effect and the u,d- quark mass difference effect.
In fact, 
if we consider the SU(2) rotation of $\exp[-i\pi\sigma^2/2]$
(i.e., u,d exchange), the $\Sigma$ triplets rotate as
$\Sigp\rightarrow\Sigm$,
$\Sigm\rightarrow\Sigp$,
$\Sigz\rightarrow\Sigz$.
Therefore, in the splitting of $\mSigall$,
there is no ${\cal O}(m_u-m_d)$ term 
and there exists only ${\cal O}(\alem)$ EM effect
up to the first order of the isospin breaking effect.
Under this consideration, 
the result from the EM effect can be directly compared with
the experimental value of $1.5$MeV, in principle.
Unfortunately, because of the statistical fluctuation,
we obtain only the zero-consistent result 
for $\mSigall$ after the linear chiral extrapolation.
The upper bound of the lattice result for this quantity is found to be
smaller than the experiental value, while 
this discrepancy may be attributed by the finite volume artifact\cite{duncan}.

%
%

Before closing this section,
we comment on the systematic error in these results.
In order to check the stability of the analysis,
we examine the several alternative methods to 
extract the mass difference.
For example, we perform the linear fit 
for the ratio of proton to neutron correlation functions 
(i.e., for the ratio shown in Fig.\ref{fig:PNratio}) in terms of $t$,
instead of exponential fit of each correlation function.
We find that the result is consistent with each other
and confirm that the results are reliable.
In the variational method, we choose several $t_0$ instead of $t_0=1$ choice
when diagonalizing $\Pi(t=t_0)^{-1} \cdot \Pi(t)$,
and check the dependence on $t_0$.
It is found that the results are insensitive to $t_0$, and 
we confirm that our variational procedure is stable.
Yet, the current lattice results are afflicted by 
the statistical noise, particularly in light quark mass sector,
and it is difficult to perform the definite chiral extrapolation.
Further calculation is desirable to achieve the better statistics,
which is actually our ongoing work.
%
%
Although the uncertainty in the  result extracted from the current lattice setup
has been evaluated as describe above,
the most troublesome artifact remained is the finite volume artifact,
because the QED interaction is a long-range interaction.
In fact, the model-based calculation\cite{duncan} 
suggests that such artifact is sometimes comparable to 
the results of the lattice simulation.
At this moment, we cannot evaluate the finite volume artifact
without using model calculations,
and the results given above would receive some modifications.
The explicit calculation of the finite volume artifact
is planned with the use of $N_f=2+1$ configurations generated by
the RBC-UKQCD collaboration\cite{rbc-2+1}, with which the configurations with
different volumes are available.

%

\section{Summary and outlook}

We have investigated the electromagnetic (EM) effect
on octet baryon spectroscopy.
By employing $N_f=2$ dynamical domain-wall QCD configuration combined with 
non-compact quenched QED configuration,
the u, d sea quark effect in QCD has been incorporated.
The mass splittings between isomultiplets
$(p, n), (\Sigp, \Sigz, \Sigm), (\Xiz, \Xim)$
have been studied
by evaluating the two-point correlation function,
where the variational method has been adopted
in the $Q=0, S=-1$ channel
in order to extract the $\Sigz$ state as a first excited state in this channel.
In order to study the QED charge dependence of the mass splitting, 
we have chosen the QED charges of 
$\alem = (0), \alem^{\rm phy}, \frac{(0.6)^2}{4\pi}, \frac{(0.85)^2}{4\pi}, \frac{(1.0)^2}{4\pi}$.
We have found that it is essential 
to calculate both of $e = \pm\sqrt{4\pi\alem}$ for each $\alem$,
in order to cancel the ${\cal O}(e)$ contamination and 
to achieve the reasonable S/N in the baryon EM mass splitting.
By fitting the mass splitting linearly in terms of $\alem$,
we have obtained the first result from lattice dynamical simulation
for the baryon EM mass splitting at each $m=0.02, 0.03, 0.04$.
The investigation of the effect of 
u, d quark mass difference is also in progress\cite{doi-future}.
There remains uncertainty originates from the finite volume artifact.
In order to investigate it explicitly,
we are planning to perform the analysis
using the $N_f=2+1$ dynamical domain-wall configuration
generated by the RBC-UKQCD collaboration\cite{rbc-2+1}.
There, we can also eliminate the quenching artifact 
for the strange quark as well.
In future, the inclusion of the dynamical QED effect
is interesting to investigate\cite{duncan-full-qed},
in which we expect the definite calculation of the 
isospin breaking effect is possible
with all the uncertainties in the simulation under control.

\acknowledgments
We thank Mr. Ran Zhou for his useful help.
We thank RIKEN, Brookhaven National Laboratory and 
the U.S. Department of Energy for providing the facilities 
essential for the completion of this work.
T.~D. is supported by Special Postdoctoral Research Program of RIKEN.
This work is supported in part by the Grant-in-Aid of the Ministry of
Education (No. 18034011, 18340075, 18740167)
and
DOE Outstanding Junior Investigator grant No. DEFG0292ER40716.

\end{document}